\documentclass[twocolumn,aps,superscriptaddress,showpacs,nofootinbib]{revtex4}
\usepackage{hyperref}
\usepackage{footnote}
\usepackage{amssymb}
\usepackage{amsmath}
\usepackage{graphicx}
\usepackage[normalem]{ulem}
\usepackage{url}
\usepackage{csquotes}
\usepackage{color}

\begin{document}



\title{Isospin effect on the liquid-gas phase transition for finite nuclei}

\author{S. Mallik}
\email{swagato@vecc.gov.in}
\affiliation{Physics Group, Variable Energy Cyclotron Centre, 1-AF Bidhan Nagar, Kolkata 700064, India}
\affiliation{Homi Bhabha National Institute, Training School Complex, Anushakti Nagar, Mumbai 400085, India}

%\affiliation[first]{organization={University of the Moon},%Department and Organization
%            city={Earth},

\begin{abstract}
The phenomenon of nuclear liquid–gas phase transition is a topic of contemporary interest. In heavy-ion collisions, there is no direct way of accessing the thermodynamic variables like pressure, density, free energy, entropy etc., and unambiguous detection of phase transition becomes difficult. A peak in the first order derivative of total multiplicity with respect to temperature (commonly abbreviated as the multiplicity derivative) has been established as a new experimentally accessible signature of the nuclear liquid-gas phase transition. In this work, the effect of isospin asymmetry in the fragmenting system, as well as the nuclear equation of state, on the multiplicity derivative and specific heat at constant volume is investigated within the framework of the Canonical Thermodynamical Model (CTM) with a semi-microscopic cluster functional.
\end{abstract}

%\begin{keyword}
%Nuclear liquid-gas phase transition \sep Nuclear multifragmentation \sep Intermediate energy heavy-ion reactions \sep Equation of state \sep Multiplicity derivative
%\end{keyword}

%\tableofcontents


\maketitle
\section{Introduction}
Phase transition is a thermodynamic process where a system changes from one phase or state to another by transfer of energy \cite{Stanley}. The Lenard-Jones potential for molecular interaction, is repulsive at very short range and then at comparatively higher intermolecular separation it is attractive. Turning to nuclear physics, the nuclear equation of state provides a way to describe the bulk properties of a nuclear many body system in thermodynamical equilibrium, governed by the nucleon-nucleon interaction at the microscopic level. If one studies the nucleon-nucleon interaction potential, it is observed that its variation with separating distance is similar to the Lenard-Jones potential (though the scales of interaction strength and range are widely different) \cite{Mallik_book_chapter}. This introduces the concept of nuclear liquid-gas phase transition. For the last few decades, one of the primary motivations of nuclear physics community is to probe the liquid-gas coexistence region in the phase diagram of nuclear matter \cite{Borderie,Borderie2,DasGupta_book,Siemens,Gross_phase_transition,Bondorf1,Dasgupta_Phase_transition,Chomaz,Chomaz2}. Nuclear multifragmentation is a unique method for studying nuclear phase transitions, as it covers the entire fragment spectrum. However, due to the finite size of atomic nuclei and the presence of long-range Coulomb interactions, it is extremely difficult to establish such a transition. Consequently, a series of both theoretical and experimental studies have been conducted to investigate the nuclear liquid-gas phase transition, particularly through central collision reactions at Fermi energies and projectile fragmentation reactions at relativistic energies.\\
\indent
In heavy-ion reactions there is no direct way to measure the thermodynamic variables like pressure, density, free energy, entropy etc. Therefore different signals of nuclear phase transition which have been explored so far are the caloric curve \cite{Pochodzalla_phase_transition}, the negative heat capacity \cite{Agostino1,Agostino2}, bimodality in charge asymmetry \cite{Chomaz_bimo,Gulminelli1,Krishnamachari,Pleimling_JPA,Chaudhuri_largest_cluster,Fevre1,Mallik14}, Landau free energy approach \cite{Huang_LPT,Bonasera_LPT}, spinodal decomposition \cite{Heiselberg,Lopez,Colonna}, fluctuation properties of the largest cluster \cite{Borderie_JPG,Dorso}, the moment of the charge distributions \cite{Campi}, Fisher’s power-law exponent of fragments \cite{Elliott} and Zipf's law \cite{Ma} etc.  To search more direct signature of phase transition the variation of first order derivative of total fragment multiplicity with respect to temperature (commonly abbreviated as multiplicity derivative) has been proposed \cite{Mallik16}. First theoretical work on multiplicity derivative \cite{Mallik16} has been performed in the framework of Canonical Thermodynamical model (CTM) \cite{Das} of nuclear multifragmentation. This work concluded that the temperature corresponding to the peak of the multiplicity derivative and specific heat at constant volume appears at same temperature and the nuclear phase transition is first order. Later this proposed signal of multiplicity derivative with respect to temperature was tested and verified in different statistical and dynamical models like the statistical multifragmentation model (SMM) \cite{Lin1,Lin2}, Quantum Molecular Dynamics (QMD) model \cite{Liu}, lattice gas model \cite{Mallik20}, percolation model \cite{Mallik20}, static antisymmetrized molecular dynamics model \cite{Lin3} and Nuclear statistical Equilibrium (NSE) model \cite{Bakeer}. The theoretical proposition of this signal got further support when it was experimentally verified by measuring the quasiprojectile reconstructed from the reactions $^{40}Ar + ^{58}Ni$ , $^{40}Ar + ^{27}Al$  and $^{40}Ar + ^{48}Ti$ at 47 Mev/nucleon performed at Texas A$\&$M university $K$=500 superconducting cyclotron \cite{Wada}.\\
\indent
The properties of nuclear liquid gas phase transition are correlated to the nuclear equation of state (EoS) at sub-saturation densities and finite temperatures \cite{Bao-an-li2,Bao-an-li1}. A detailed knowledge of the EoS (i.e. the dependence of the pressure or alternatively, of the energy per nucleon on the temperature and the density) is essential to describe different aspects of nuclear physics as well as nuclear astrophysics. Recently, the CTM has been upgraded with semi-microscopic cluster functional \cite{Mallik25} where the bulk part of the binding and excitation have been determined from the meta-modelling of the equation of state \cite{Margueron2018} with Sly5 parameters \cite{Sly5}. Based on this, the aim of this present work, is to study how this cluster functional affects the signatures of phase transition like multiplicity derivative with respect to temperature and specific heat. In addition to that, the bahavior of these signatures due to different realistic nuclear EoS, mass and isospin asymmetry of the fragmenting system will be examined.\\
\indent
The paper is structured as follows. In section 2, a brief introduction of the Canonical Thermodynamical model is presented. The results are described in section 3, finally summary is discussed in section 4.
\section{Model Description}
In CTM, \cite{DasGupta_book,Das} it is assumed that statistical equilibrium is attained at freeze-out stage. Population of different channels of disintegration is solely decided by statistical weights in the available phase space. The calculation is done for a fixed mass and atomic number, freeze out volume and temperature. In a canonical model \cite{Das}, the partitioning is done such that all partitions have the correct $A_{0},Z_{0}$ (equivalently $N_{0},Z_{0}$). The canonical partition function is given by
\begin{eqnarray}
Q_{N_{0},Z_{0}} & = & \sum\prod\frac{\omega_{N,Z}^{n_{N,Z}}}{n_{N,Z}!}
\end{eqnarray}
where the sum is over all possible channels of break-up (the number of such channels is enormous) satisfying $N_{0}=\sum N\times n_{N,Z}$ and $Z_{0}=\sum Z\times n_{N,Z}$; $\omega_{N,Z}$ is the partition function of the composite with $N$ neutrons \& $Z$ protons and $n_{NZ}$ is its multiplicity. The partition function $Q_{N_{0},Z_{0}}$ is calculated by applying a recursion relation \cite{Chase}. From Eq. (1), the average number of composites with $N$ neutrons and $Z$ protons can be expressed as,
\begin{eqnarray}
\langle n_{N,Z}\rangle & = & \omega_{N,Z}\frac{Q_{N_{0}-N,Z_{0}-Z}}{Q_{N_{0},Z_{0}}}
\end{eqnarray}
\indent
The partition function of a composite having $N$ neutrons and $Z$ protons is a product of two parts: one is due to the the translational motion and the other is the intrinsic partition function of the composite:
\begin{eqnarray}
\omega_{N,Z}=\frac{V}{h^{3}}(2\pi mT)^{3/2}A^{3/2}\times z_{N,Z}(int)
\end{eqnarray}
Here, $V$ is the volume available to the clusters and free nucleons for translational motion. Assuming that the clusters can not overlap, the available volume can be expressed as $V=V_{f}-V_{ex}$ , where  $V_{f}$ is the freeze-out volume and $V_{ex}$ is the excluded volume, which is considered as constant and equal to the normal nuclear volume of fragmenting system with $Z_{0}$ protons and $N_{0}$ neutrons \cite{Bondorf1,Das}. In this work the freeze-out volume is kept constant at $6V_0$ as fragmentation from central collision reactions at intermediate energies are considered. $z_{N,Z}(int)$ is the internal partition function, the proton and the neutron are fundamental building blocks, thus $z_{1,0}(int)=z_{0,1}(int)=2$, where 2 takes care of the spin degeneracy. For $^2$H, $^3$H, $^3$He, $^4$He, $^5$He and $^6$He, $z_{N,Z}(int)=(2s_{N,Z}+1)\exp[-E_{N,Z}(gr)/T]$ where $E_{N,Z}(gr)$ is the ground-state energy of the composite and $(2s_{N,Z}+1)$ is the experimental spin degeneracy of the ground state. Excited states for these very low-mass nuclei are not included (as the excitation energies of these excited states are too high).\\
\indent
However, intermediate energy heavy-ion reactions occur at finite temperatures, and the produced clusters are no longer in their ground state. The free energy is a more comprehensive quantity that includes both energy (binding as well as excitation) and entropy contributions. So, the internal partition function of clusters with $Z \geq$3 can be expressed as,
\begin{eqnarray}
z_{N,Z}(int)&=&\exp^{-\frac{F_{N,Z}}{T}}\nonumber\\
&=&\exp^{-\frac{B_{N,Z}+E^{*}_{N,Z}-TS_{N,Z}}{T}}
\label{Free_energy}
\end{eqnarray}
where, $B_{N,Z}$, $E^{*}_{N,Z}$ and $S_{N,Z}$  are the ground state binding energy, excitation and entropy respectively, of the fragment with $Z$ protons and  $N$ neutrons. In conventional CTM approach as well as in most of the other statistical models of multifragmentation, for determining the internal partition function of nuclei with $Z \geq$3 the liquid-drop formula is used for calculating the binding energy and the remaining part of free energy is taken from the Fermi-gas model. In a recent work \cite{Mallik25}, the internal partition function of the CTM is upgraded by the semi-microscopic approach where the Helmholtz free energy of a nucleus with $N$ neutrons and $Z$ protons can be decomposed as \cite{Mallik_NuclAstro1},
\begin{equation}
F_{N,Z}=F^{bulk}+F^{surf}+F^{coul} . \label{eq:leptodermous}
\end{equation}
where the bulk part $F^{bulk}$ is originated from bulk nuclear matter at baryonic density $\rho_c=\rho_{c,n}+\rho_{c,z}$ ($\rho_{c,n}$ and $\rho_{c,z}$ are neutron and proton density respectively) and isospin asymmetry $\delta_c= (\rho_{c,n}-\rho_{c,z})/\rho_c=\frac{N-Z}{N+Z}$ occupying a finite spatial volume $V_c=(N+Z)/\rho_c$. The baryonic density with isospin asymmetry $\delta_c$ is approximated \cite{Gulminelli2015}to the corresponding saturation density ($\rho_0$) of symmetric nuclear matter at finite asymmetry according to:
\begin{eqnarray}
\rho_c(\delta_c)=\rho_0\bigg{(}1-\frac{3L_{sym}\delta_c^2}{K_{sat}+K_{sym}\delta_c^2}\bigg{)}.
\label{baryon_density}
\end{eqnarray}
The above expression is obtained from the assumption of first order derivative of the binding with respect to baryonic density equal to zero  \cite{papa} and consideration up to the curvature term in generalized liquid drop approach \cite{Ducoin1}.
The bulk part of the Helmholtz free energy given by,
\begin{eqnarray}
F^{bulk}=V_c\bigg{[}-\frac{2}{3}\sum_{q=n,z}\xi_{c,q}+\sum_{q=n,z}\rho_{c,q}\eta_q+v(\rho_c,\delta_c)\bigg{]}
\end{eqnarray}
where $\xi_{c,n}$ and $\xi_{c,z}$ are the kinetic energy density of the nucleus due to neutron ($q=n$) and proton ($q=p$) contribution respectively, which can be expressed as $q=n,p$,
\begin{eqnarray}
\xi_{c,q}&=& \frac{3h^2}{2\pi m^*_{c,q}}\bigg{(}\frac{2\pi m^*_{c,q}T}{h^2}\bigg{)}^{5/2}F_{3/2}(\eta_{c,q})
\end{eqnarray}
with
$\eta_{c,q}=F^{-1}_{1/2}\bigg{\{}\bigg{(}\frac{2\pi m^*_{g,q}T}{h^2}\bigg{)}^{3/2}\rho_{c,q}\bigg{\}}$, $F_{1/2}$ and $F_{3/2}$ are the Fermi integrals. The expression of potential energy per particle that can be adapted to different effective interactions and energy functionals  is given by:
\begin{eqnarray}
v(\rho_c,\delta_c)&=&\sum_{k=0}^{N}\frac{1}{k!}(v^{is}_{k}+v^{iv}_{k}\delta_c^2)x^{k}\nonumber\\
&+&(a^{is}+a^{iv}\delta_c^2)x^{N+1}\exp(-b\frac{\rho_c}{\rho_0}) ,
\label{ELFc_potential}
\end{eqnarray}
where $x=\frac{\rho_c-\rho_0}{3\rho_0}$, $a^{is}=-\sum_{k\geq0}^{N}\frac{1}{k!}v^{is}_{k}(-3)^{N+1-k}$ and $a^{iv}=-\sum_{k\geq0}^{N}\frac{1}{k!}v^{iv}_{k}(-3)^{N+1-k}$. $N=4$ and $b=10ln2$ are chosen for this model. This value of $b$ leads to a good reproduction of the Sly5 functional which is used for the numerical applications presented in this paper. The model parameters $v_k^{is(iv)}$ can be linked with a one-to-one correspondence to the usual EoS empirical parameters \cite{Margueron2018}, via:
\begin{eqnarray}
v^{is}_{0}&=&E_{sat}-t_0(1+\kappa_0)\nonumber\\
v^{is}_{1}&=&-t_0(2+5\kappa_0)\nonumber\\
v^{is}_{2}&=&K_{sat}-2t_0(-1+5\kappa_0)\nonumber\\
v^{is}_{3}&=&Q_{sat}-2t_0(4-5\kappa_0)\nonumber\\
v^{is}_{4}&=&Z_{sat}-8t_0(-7+5\kappa_0)
\label{Isoscalar_parameters}
\end{eqnarray}
\begin{eqnarray}
v^{iv}_{0}&=&E_{sym}-\frac{5}{9}t_0[(1+(\kappa_0+3\kappa_{sym})]\nonumber\\
v^{iv}_{1}&=&L_{sym}-\frac{5}{9}t_0[(2+5(\kappa_0+3\kappa_{sym})]\nonumber\\
v^{iv}_{2}&=&K_{sym}-\frac{10}{9}t_0[(-1+5(\kappa_0+3\kappa_{sym})]\nonumber\\
v^{iv}_{3}&=&Q_{sym}-\frac{10}{9}t_0[(4-5(\kappa_0+3\kappa_{sym})]\nonumber\\
v^{iv}_{4}&=&Z_{sym}-\frac{40}{9}t_0[(-7+5(\kappa_0+3\kappa_{sym})] \ ,
\label{Isovector_parameters}
\end{eqnarray}
\begin{figure}[!b]
\begin{center}
\includegraphics[width=\columnwidth]{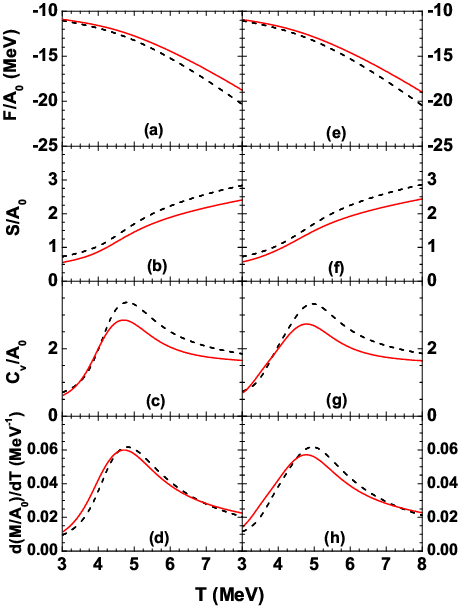}
\caption{Variation of free energy per nucleon (upper panels), entropy per nucleon (upper middle panels), specific heat per particle (lower middle panel) and multiplicity derivative (lower panel) with temperature studied from conventional CTM calculation (black dashed lines) and CTM calculation with realistic Sly5 EoS (red solid lines). Left and right panels represent the result for fragmenting system $A_0$=72, $Z_0$=30 and $A_0$=186, $Z_0$=75 respectively.}
\label{Cluster_functional_comparison}
\end{center}
\end{figure}
where  $E_{sat}$, $K_{sat}$, $Q_{sat}$ and $Z_{sat}$ are saturation energy, incompressibility modulus,  isospin symmetric skewness and  kurtosis respectively and $E_{sym}$, $L_{sym}$, $K_{sym}$, $Q_{sym}$ and $Z_{sym}$ are symmetry energy, slope, and associated incompressibility,  skewness and  kurtosis respectively.
Concerning the $\kappa_0$ and $\kappa_{sym}$, they govern the density dependence of the neutron and proton effective mass according to:
\begin{equation}
\frac{m_q}{m^*_{q}(\rho_c,\delta_c)}=1+(\kappa_0 \pm \kappa_{sym}\delta)\frac{\rho_c}{\rho_0},
\label{effective_mass}
\end{equation}
with $q=n,z$. Therefore the dependence on the equation of state on this semi-microscopic model is introduced via equations from \ref{baryon_density} to \ref{effective_mass}.\\
The finite size corrections are included by the surface part of the Helmholtz free energy ($F^{surf}$) \cite{papa} for which we adopt the prescription proposed in ref.\cite{LS,Carreau2019,Carreau2019a} on the basis of Thomas-Fermi calculations with extreme isospin ratios:
\begin{eqnarray}
F^{surf}&=&4\pi r^2_0 A^{2/3}\sigma(y_{c,p},T)
\label{Surface}
\end{eqnarray}
with $r_0=\bigg{\{}\frac{3}{4\pi\rho_0}\bigg{\}}^{1/3}$, $y_{c,p}=Z/(Z+N)$ and
\begin{eqnarray}
\sigma(y_{c,p})=\sigma_0 h\bigg{(}\frac{T}{T_c(y_{c,p})}\bigg{)}\frac{2^{p+1}+b_s}{y_{c,p}^{-p}+b_s+(1-y_{c,p})^{-p}}
\end{eqnarray}
where $\sigma_0$ represents the surface tension of symmetric nuclear matter and $b_s$ and $p$ represent the isospin dependence. For this work, $\sigma_0=1.09191$, $b_s=15.36563$ and $p=3.0$ are used. One could also incorporate the effect of the nuclear equation of state (EoS) on the surface tension by determining the surface energy parameters using a Bayesian approach. However, in this work, we focus on studying the EoS effect on the bulk part only and the rigorous determination of EoS-dependent surface energy parameters is beyond the scope of this present study. The Coulomb contribution in Helmholtz free energy is considered as same as before i.e.
\begin{eqnarray}
F^{Coul}&=&a^{*}_c\frac{Z^2}{A^{1/3}}
\label{Coulomb}
\end{eqnarray}
[$a^{*}_c=0.31a_{c}$ with $a_{c}=0.72$ MeV and Wigner-Seitz correction factor 0.31 \cite{Bondorf1}]. A detailed expression for the nuclear binding energy, $B_{N,Z}$ (as mentioned in Eq. \ref{Free_energy}) at zero temperature is provided in Appendix.
\begin{figure}[!b]
\begin{center}
\includegraphics[width=\columnwidth]{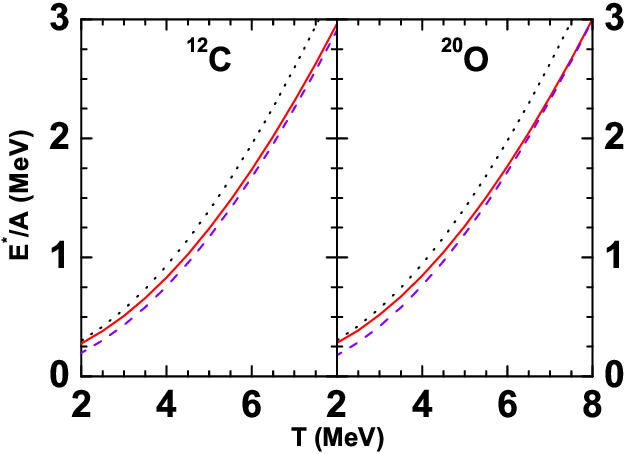}
\caption{Excitation of $^{12}$C (left panel) and $^{20}$O (right panel) determined by using the Sly5 (red solid lines), SGII (black dotted lines) and NL3 (violet dashed lines) equation of state.}
\label{Excitation_figure}
\end{center}
\end{figure}
\section{Results}
Effect of changing the cluster functional from semi-empirical binding (determined from Bethe-Weizsacker mass formula) and excitation (determined from Fermi gas model) to semi-microscopic realistic nuclear EoS on the nuclear phase transition signatures is presented in Fig. \ref{Cluster_functional_comparison}. Sly5 EoS parameters \cite{Sly5} are used for determining the semi-microscopic cluster functional calculation \cite{Mallik25}. Disintegration of two systems having (i) mass number $A_0$=72, atomic number $Z_0$=30 and (ii) $A_0$=186, $Z_0$=75 are simulated which are expected to be formed in central collision of $^{48}$Ca+$^{48}$Ca and $^{124}$Sn+$^{124}$Sn reactions respectively with $25\%$ pre-equilibrium emission \cite{Frankland,Xu, Mallik11,Mallik23}. For both kind of cluster functional, free energy per nucleon is continuous against $T$ and due to presence of long range Coulomb interaction sudden increase in entropy (which occurs in first order phase transition for finite nuclei as described in Fig. 1 of Ref. \cite{Mallik19}) is not clearly visible. However, specific heat per particle at constant volume ($C_v/A_0$) and multiplicity derivative normalised by fragmenting system mass number $\frac{d(M/A_0)}{dT}$ shows a peak for both cases. For each type cluster functional, the temperatures at which $C_v/A_0$ and $\frac{d(M/A_0)}{dT}$ are maximum are almost identical. This interesting aspect further motivates to study the effect of different nuclear EoS on these signatures of nuclear phase transition in the framework of CTM model.\\
\begin{table}[!t]
\begin{center}
\begin{tabular}{|p{1.85cm}|p{1.45cm}|p{1.45cm}|p{1.45cm}|}
\hline
Nuclear EoS & Sly5 & SGII & NL3 \\
\hline
$\rho_0$ (fm$^{-3})$ &0.1604 & 0.1583 & 0.1480\\
$E_{sat}$ (MeV) & -15.98 & -15.59 & -16.24\\
$E_{sym}$ (MeV) & 32.03 & 26.83 & 37.35\\
$L_{sym}$ (MeV) & 48.3 & 37.6 & 118.3\\
$K_{sat}$ (MeV) & 230 & 215 & 271\\
$K_{sym}$ (MeV) & -112 & -146 & 101\\
$Q_{sat}$ (MeV) & -364 & -381 & 198\\
$Q_{sym}$ (MeV) & 501 & 330 & 182\\
$Z_{sat}$ (MeV) & 1592 & 1742 & 9302\\
$Z_{sym}$ (MeV) & -3087 & -1891 & -3961\\
$\kappa_{sat}$  & 0.43 & 0.27 & 0.49\\
$\kappa_{sym}$ & 0.18 & -0.22 & -0.09\\
\hline
\end{tabular}
\end{center}
\caption{Set of nuclear equation of state parameters used for determining cluster functional.}
\label{table_parameter}
\end{table}
\begin{figure}[!b]
\begin{center}
\includegraphics[width=\columnwidth]{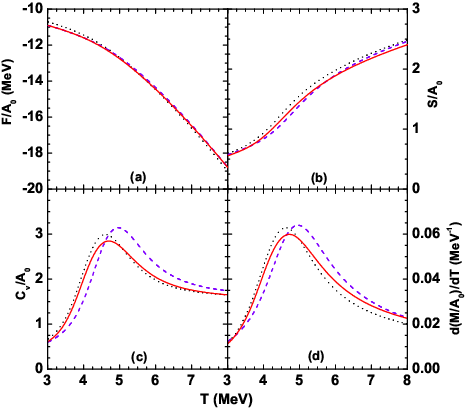}
\caption{Variation of free energy per nucleon (upper left panel), entropy per nucleon (upper right panel), specific heat per particle (lower left panel) and multiplicity derivative (lower right panel) with temperature studied from CTM calculation with realistic Sly5 (red solid lines), SGII (black dotted lines) and NL3 (violet dashed lines) for the fragmenting system having $A_0$=72 and $Z_0$=30.}
\label{EoS_comparison_Z0=30_A0=72}
\end{center}
\end{figure}
\begin{figure}[!t]
\begin{center}
\includegraphics[width=\columnwidth]{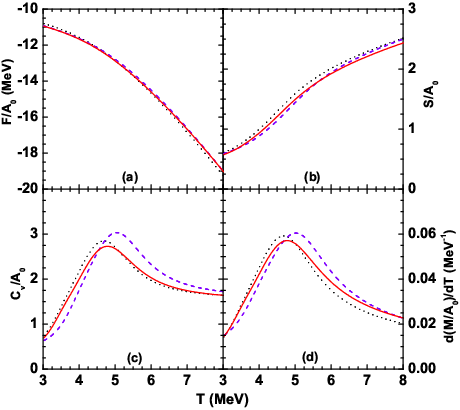}
\caption{Same as Fig. \ref{EoS_comparison_Z0=30_A0=72} except that here the mass and atomic number of the fragmenting system are $A_0$=186 and $Z_0$=75 respectively.}
\label{EoS_comparison_Z0=75_A0=186}
\end{center}
\end{figure}

\begin{figure}[!b]
\begin{center}
\includegraphics[width=\columnwidth]{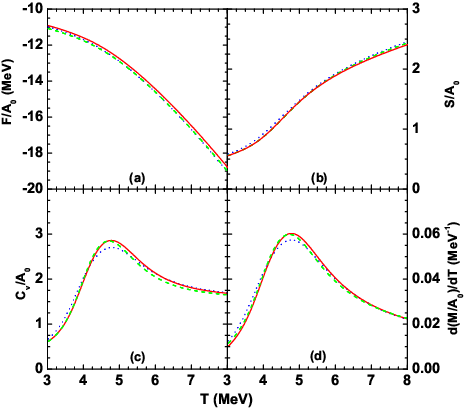}
\caption{Variation of free energy per nucleon (upper left panel), entropy per nucleon (upper right panel), specific heat per particle (lower left panel) and multiplicity derivative (lower right panel) with temperature studied for the three fragmenting system having same $Z_0$=30 but different $A_0$=60 (blue dotted lines), 66 (green dashed lines) and 72 (red solid lines). CTM calculations are performed by determining the cluster functional from Sly5 EoS parameters.}
\label{Isospin_comparison_Z0=30}
\end{center}
\end{figure}
\indent
In order to do that, in addition to Sly5 EoS, the semi-microscopic cluster functional of the CTM described in Section II is determined from NL3 \cite{Lalazissis} and SGII \cite{Nguyen} EoS also. Isoscalar and isovector parameters used in this paper for three different EoS are mentioned in Table-\ref{table_parameter}. Nuclear phase transition from heavy-ion reactions at intermediate energies can be explained as the competition between the surface energy and excitation energy. The surface tension effect tries to accumulate more nucleons together (i.e. prefers liquid phase) on the other side excitation of the nucleus tries to break it in to free nucleons and small composites (i.e. prefers gas phase). Here surface energy of a fragment is kept identical (as described in the last part of Section-II), but the excitation energy of different fragments at a given temperature obtained from SGII (NL3) EoS are higher (lower) compared to that of Sly5 EoS. This bulk excitation energy per nucleon of the fragment with $Z$ protons and $N$ neutrons (at baryonic density $\rho_c$ and isospin asymmetry $\delta_c$) is determined from the expression,
\begin{eqnarray}
\frac{E^*_{N,Z}(\rho_c,T)}{A}&=&\frac{1}{{\rho_c}}\sum_{q=n,z}\bigg{[}\frac{3h^2}{2\pi m^*_{c,q}}\bigg{(}\frac{2\pi m^*_{c,q}T}{h^2}\bigg{)}^{5/2}F_{3/2}(\eta_{c,q})\bigg{]}\nonumber\\
&-&\frac{t_0}{2}\bigg{(}\frac{\rho_c}{\rho_0}\bigg{)}^{2/3}\bigg{[}(1+\kappa_0\frac{\rho_c}{\rho_0})f_1(\delta_c)\nonumber\\
&+&\kappa_{sym}\frac{\rho_c}{\rho_0}f_2(\delta_c)\bigg{]}
\end{eqnarray}
where, $t_0=\frac{3\hbar^2}{10m}\big{(}\frac{3\pi^2 \rho_0}{2}\big{)}^{2/3}$ is kinetic energy per nucleon in symmetric matter at saturation and the functions $f_1(\delta_c)=(1+\delta_c)^{5/3}+(1-\delta_c)^{5/3}$ and $f_2(\delta_c)=\delta_c[(1+\delta_c)^{5/3}-(1-\delta_c)^{5/3}]$ \cite{Margueron2018,Mallik_NuclAstro1}. For example, the temperature dependence of excitation energy per nucleon of $^{12}$C and $^{20}$O fragments for these three considered EoS are shown in Fig. \ref{Excitation_figure}. Hence, due this change in excitation, the peak of the specific heat per particle at constant volume ($C_v/A_0$) as well as multiplicity derivative normalised by fragmenting system mass number $\frac{d(M/A_0)}{dT}$ (which represents the temperature at which nuclear liquid-gas phase transition occurs) obtained from CTM calculation with SGII (NL3) EoS shifted towards lower (higher) temperature side compare to that of Sly5 EoS. This is displayed in Fig. \ref{EoS_comparison_Z0=30_A0=72} and \ref{EoS_comparison_Z0=75_A0=186} for the two fragmenting systems $A_0$=72, $Z_0$=30 and $A_0$=186, $Z_0$=75 respectively.\\
\begin{figure}[!t]
\begin{center}
\includegraphics[width=\columnwidth]{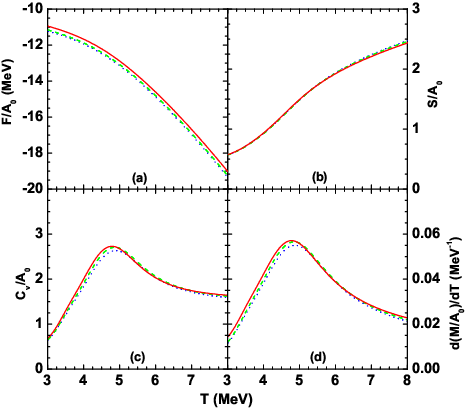}
\caption{Same as Fig. \ref{Isospin_comparison_Z0=30} except the three fragmenting systems having same $Z_0$=75 but different $A_0$=168 (blue dotted lines), 177 (green dashed lines) and 186 (red solid lines).}
\label{Isospin_comparison_Z0=75}
\end{center}
\end{figure}
\begin{figure}[!b]
\begin{center}
\includegraphics[width=\columnwidth]{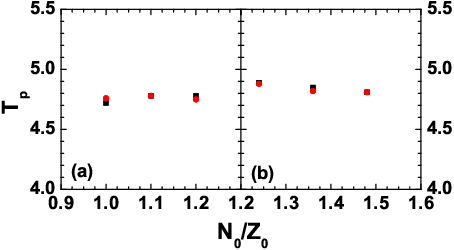}
\caption{Isospin dependence of the phase transition temperature obtained from peak of specific heat per particle (black squares) and multiplicity derivative (red circles) for two sets of fragmenting systems having atomic number $Z_0$=30 (left panel) and 75 (right panel).}
\label{Phase_transition_temperature}
\end{center}
\end{figure}
\begin{figure}[!t]
\begin{center}
\includegraphics[width=0.7\columnwidth]{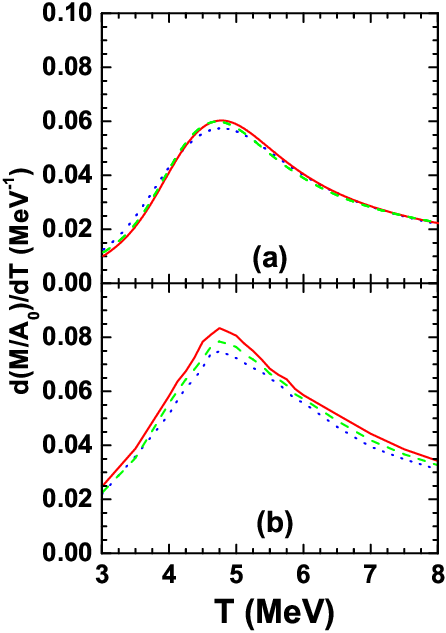}
\caption{Effect of secondary decay on multiplicity derivative for the three fragmenting systems having same $Z_0$=30 but different $A_0$=60 (blue dotted lines), 66 (green dashed lines) and 72 (red solid lines). Upper panel show the results after the multifragmentation stage (same as Fig. 5(d)) where as lower panel represent the results after secondary decay of the excited fragments.}
\label{Secondary_decay}
\end{center}
\end{figure}

\begin{figure}[!b]
\begin{center}
\includegraphics[width=\columnwidth]{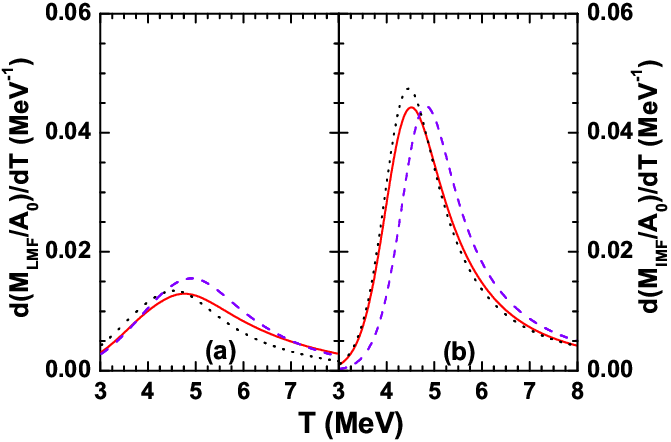}
\caption{Variation of the first order derivative of light fragment (with atomic number $Z$=$1$ and $2$) multiplicity (left column) and intermediate mass fragment (with atomic number $3{\le}Z{\le}20$) multiplicity with respect to temperature (right panel) with temperature studied from CTM calculation with realistic Sly5 (red solid lines), SGII (black dotted lines) and NL3 (violet dashed lines) for the fragmenting system of mass number $A_0$=72 and atomic number $Z_0$=30.}
\label{EoS_comparison_IMF_Z0=30_A0=72}
\end{center}
\end{figure}
\indent
The liquid gas phase transition of nuclear matter is connected to the isospin asymmetry \cite{Ducoin}. To study the effect of isospin asymmetry on these nuclear phase transition signatures for finite nuclei, CTM calculation with semi-microscopic cluster functional (with Sly5 EoS only) is performed for two sets of disintegrating system-(i) fixed $Z_0$=30 but three different $A_0$=72, 66 and 60 which are expected to be formed in central collision of $^{48}$Ca+$^{48}$Ca, $^{40}$Ca+$^{48}$Ca and $^{40}$Ca+$^{40}$Ca reactions respectively (by assuming $25\%$ pre-equilibrium emission in each reaction) and (ii) fixed $Z_0$=75 but three different $A_0$=186, 177 and 168 which are expected to be formed in central collision of $^{124}$Sn+$^{124}$Sn, $^{112}$Sn+$^{124}$Sn and $^{112}$Sn+$^{112}$Sn respectively (by assuming $25\%$ pre-equilibrium emission in each reaction). Results for set (i) and (ii) are displayed in Fig. \ref{Isospin_comparison_Z0=30} and \ref{Isospin_comparison_Z0=75} respectively. For both sets of disintegrating systems, peak of both $C_v/A_0$ and $\frac{d(M/A_0)}{dT}$ is almost independent of isospin asymmetry. The dependence of phase transition temperature with isospin asymmetry (neutron to proton ratio of the fragmenting system) obtained from both $C_v/A_0$ and $\frac{d(M/A_0)}{dT}$ is displayed in Fig. \ref{Phase_transition_temperature}. Fig. \ref{Phase_transition_temperature} concludes that the dependence of transition temperature on isospin asymmetry is very less in heavy ion reactions with  finite nuclei. \\

\begin{figure}[!t]
\begin{center}
\includegraphics[width=\columnwidth]{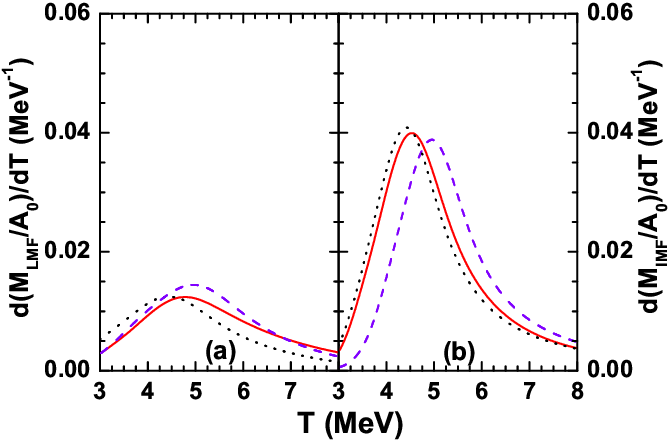}
\caption{Same as Fig. \ref{EoS_comparison_IMF_Z0=30_A0=72} except that here the mass and atomic number of the fragmenting system are $A_0$=186 and $Z_0$=75 respectively.}
\label{EoS_comparison_IMF_Z0=75_A0=186}
\end{center}
\end{figure}

\begin{figure}[!b]
\begin{center}
\includegraphics[width=\columnwidth]{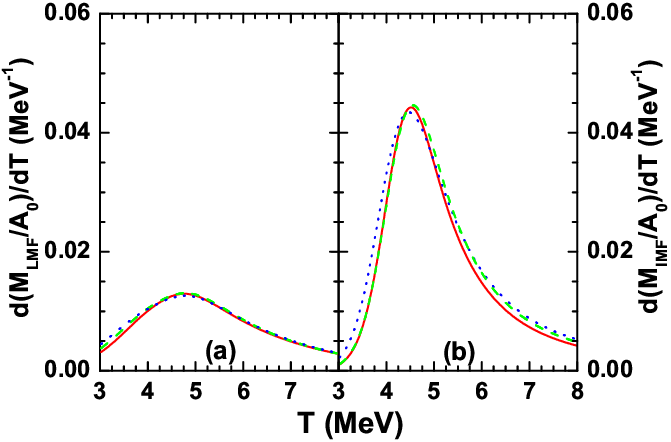}
\caption{Temperature dependence of the first order derivative of light fragment (with atomic number $Z$=$1$ and $2$) multiplicity (left column) and intermediate mass fragment (with atomic number $3{\le}Z{\le}20$) multiplicity (right panel) for the three fragmenting systems, each with the same $Z_0$=30 but different $A_0$=60 (blue dotted lines), 66 (green dashed lines) and 72 (red solid lines). CTM calculations are performed by determining the cluster functional from Sly5 EoS parameters.}
\label{Isospin_comparison_IMF_Z0=30}
\end{center}
\end{figure}

\begin{figure}[!t]
\begin{center}
\includegraphics[width=\columnwidth]{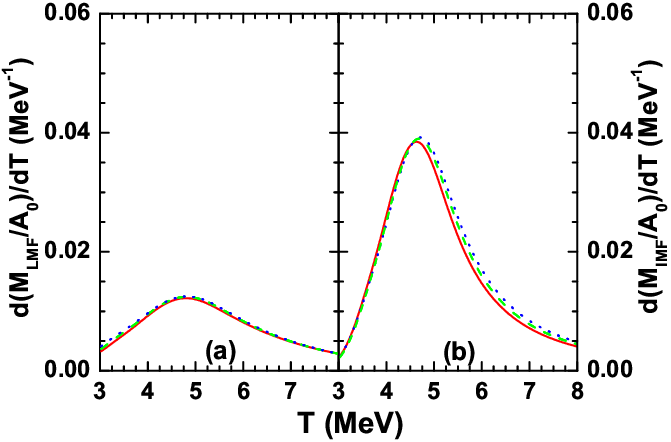}
\caption{Same as Fig. \ref{Isospin_comparison_IMF_Z0=30} but the three fragmenting systems having same $Z_0$=75 but different $A_0$=168 (blue dotted lines), 177 (green dashed lines) and 186 (red solid lines).}
\label{Isospin_comparison_IMF_Z0=75}
\end{center}
\end{figure}
\indent
The excited fragments produced in the multifragmentation stage decay to their stable ground states before reaching the detector. Hence in experiments, total multiplicity will be modified. Effect of secondary decay on experimentally accessible signature $\frac{d(M/A_0)}{dT}$ is examined for fixed $Z_0$=30 but three different $A_0$=72, 66 and 60. This is presented in Fig. \ref{Secondary_decay}. For each fragmenting systems different isospin asymmetries the peak position in multiplicity derivative with respect to temperature remains unchanged even after the secondary decay and the peaks become sharper.\\
\indent
The intermediate mass fragment (with atomic number $3{\le}Z{\le}20$) multiplicity
($M_{IMF}$) is another key observable in nuclear reactions at intermediate energies, which has been studied both experimentally and theoretically in various scenarios \cite{Peaslee,Ogilvie}. The first order derivatives of the intermediate mass fragment multiplicity and light fragment (with atomic number $Z$=$1$ and $2$) multiplicity with respect to temperature ($\frac{d(M_{IMF}/A_0)}{dT}$ and $\frac{d(M_{LMF}/A_0)}{dT}$ respectively) are studied from the CTM with Sly5, NL3 and SGII EoS for the same two disintegrating systems (i) $A_0$=72, $Z_0$=30 and (ii) $A_0$=186, $Z_0$=75 (displayed in Fig. \ref{EoS_comparison_IMF_Z0=30_A0=72} and \ref{EoS_comparison_IMF_Z0=75_A0=186} respectively). For both observables, a phase transition signal is observed, and the temperatures at which the maxima of $\frac{d(M_{IMF}/A_0)}{dT}$ and $\frac{d(M_{LMF}/A_0)}{dT}$ occur are very close to the peak of $\frac{d(M/A_0)}{dT}$ and $C_v/A_0$ for each of these three EoS.\\
\indent
The effect of isospin asymmetry of the fragmenting system on $\frac{d(M_{IMF}/A_0)}{dT}$ and $\frac{d(M_{LMF}/A_0)}{dT}$ is also examined within the CTM framework using the Sly5 EoS. This analysis considers same two sets of systems as before: (i) the same $Z_0$=30 but different $A_0$=60, 66 and 72 and (ii) the same $Z_0$=75 but different $A_0$=168, 177 and 186 (displayed in Fig. \ref{Isospin_comparison_IMF_Z0=30} and \ref{Isospin_comparison_IMF_Z0=75} respectively). These two figures indicate that, similar to the specific heat and total multiplicity derivative, the peak positions of both the intermediate mass fragment and light fragment multiplicity derivatives are almost independent of isospin asymmetry.\\
\indent
\section{Summary}
The canonical thermodynamical model with realistic semi-microscopic cluster functional has been applied to study the isospin effect on liquid-gas phase transition for finite nuclei. CTM with this upgraded cluster functional also shows peak in specific heat as well as multiplicity derivative, and their behavior is almost  identical with the CTM calculation by using binding energy from liquid drop model and excitation from Fermi gas model. However, the phase transition temperature obtained from the specific heat and the first-order derivatives of total multiplicity, light fragment multiplicity, and intermediate mass fragment multiplicity with respect to temperature is sensitive to the different nuclear EoS (Sly5, SGII, and NL3). Behavior of the specific heat and different multiplicity derivative distributions with temperature are almost independent of the isospin asymmetry of the dissociating system. This isospin independence of the phase transition temperature obtained from multiplicity derivative is also present after the secondary decay of the excited fragments.
\section{Appendix: Density dependent nuclear binding of clusters}
To extend the metamodeling of infinite nuclear matter to finite nuclei and account for their surface and Coulomb properties, the binding energy can be expressed within the compressible liquid drop approximation \cite{LS,Baym}. For a cluster with proton number $Z$ and neutron number $N$, the binding energy can be expressed as:
\begin{eqnarray}
B_{N,Z}&=&B^{bulk}+B^{surf}+B^{coul}\nonumber\\
&=&A\{t(\rho_c,\delta_c,T=0)+v(\rho_c,\delta_c)\}\nonumber\\
&&+4\pi r^2_0A^{2/3}\sigma_{T=0}(y_{c,p})+a_c\frac{Z^2}{A^{1/3}}\nonumber\\
&=&A\bigg{[}\frac{t_0}{2}\bigg{(}\frac{\rho_c}{\rho_0}\bigg{)}^{2/3}\bigg{\{}(1+\kappa_0\frac{\rho_c}{\rho_0})f_1(\delta_c)
+\kappa_{sym}\frac{\rho_c}{\rho_0}f_2(\delta_c)\bigg{\}}\nonumber\\
&+&\sum_{k=0}^{N}\frac{1}{k!}(v^{is}_{k}+v^{iv}_{k}\delta_c^2)x^{k}\nonumber\\
&+&(a^{is}+a^{iv}\delta_c^2)x^{N+1}\exp(-b\frac{\rho_c}{\rho_0})\bigg{]}\nonumber\\
&+&4\pi r^2_0 A^{2/3}\sigma_0\frac{2^{p+1}+b_s}{y_{c,p}^{-p}+b_s+(1-y_{c,p})^{-p}}+a_c\frac{Z^2}{A^{1/3}}
\label{binding_formula}
\end{eqnarray}
where each term, parameter, and coefficient has already been introduced in the model description section.\\


\begin{thebibliography}{99}
\bibitem{Stanley} H.E. Stanley, {\it Introduction to phase transitions and critical phenomena}, (Oxford University Press  (1971).
\bibitem{Mallik_book_chapter} S.Mallik and G.Chaudhuri, Chapter-4, {\it Multifragmentation in Heavy-Ion Reactions}, edited by R. K. Puri, Y. G. Ma and A. Sharma, Jenny Stanford Publishing (2023).
\bibitem{Borderie} B.Borderie and J.D.Frankland, Liquid–Gas phase transition in nuclei,  Prog. Part. Nucl. Phys. {\bf 105}, 82 (2019).
\bibitem{DasGupta_book} S. Das Gupta, S. Mallik and G. Chaudhuri, {\it Heavy ion reaction at intermediate energies: Theoretical Models}, World Scientific Publishers (2019).
\bibitem{Borderie2} B. Borderie and M. F. Rivet, Prog. Part. Nucl. Phys. {\bf 61}, 551 (2008).
\bibitem{Siemens} P.J. Siemens, Nature, {\bf 305}, 410 (1983).
\bibitem{Gross_phase_transition} D. H. E. Gross, Prog. Part. Nucl. Phys. {\bf 30}, 155 (1993).
\bibitem{Bondorf1} J. P. Bondorf, A. S. Botvina, A. S. Iljinov, I. N. Mishustin and K. Sneppen, Phys. Rep. {\bf 257}, 133 (1995).
\bibitem{Dasgupta_Phase_transition} S. Das Gupta, A. Z. Mekjian and M. B. Tsang, {\it Advances in Nuclear Physics}, Vol. 26, 89 (2001) edited by J. W. Negele and E. Vogt, Plenum Publishers, New York.
\bibitem{Chomaz} P.Chomaz et. al., Phys. Rep. {\bf 389} 263 (2004).
\bibitem{Chomaz2} F. Gulminelli et al.,; Phys. Rev. Lett. {\bf 91}, 202701 (2003).
\bibitem{Pochodzalla_phase_transition} J. Pochodzalla et al., Phys. Rev. Lett. {\bf 75}, 1040 (1995).
\bibitem{Agostino1}M. D’Agostino et al., Phys. Lett. B {\bf473}, 219 (2000).
\bibitem{Agostino2}M. D’Agostino et al., Nucl. Phys. A {\bf 699}, 795 (2002).
\bibitem{Chomaz_bimo} P. Chomaz, V. Duflot, and F. Gulminelli, Phys. Rev. Lett. {\bf 85}, 3587 (2000).
\bibitem{Gulminelli1} F. Gulminelli and Ph. Chomaz, Phys. Rev. C {\bf 71}, 054607 (2005).
\bibitem{Krishnamachari} B. Krishnamachari, J. McLean, B. Cooper, and J. Sethna, Phys. Rev. B {\bf 54}, 8899 (1996).
\bibitem{Pleimling_JPA} M. Pleimling and W. Selke, J. Phys. A {\bf 33}, L199 (2000).
\bibitem{Chaudhuri_largest_cluster} G. Chaudhuri and S. Das Gupta., Phys. Rev C {\bf 75}, 034603 (2007).
\bibitem{Fevre1} A. Le Fevre and J. Aichelin, Phys. Rev. Lett. {\bf 100}, 042701 (2008)
\bibitem{Mallik14}S.Mallik, S.Das Gupta and G.Chaudhuri, Phys. Rev. {\bf C 93} 041603 (2016) (R).
\bibitem{Huang_LPT}M. Huang et al., Phys. Rev. C {\bf 82}, 054602 (2010).
\bibitem{Bonasera_LPT}A. Bonasera et al., Phys. Rev. Lett. {\bf 101}, 122702 (2008).
\bibitem{Heiselberg} H. Heiselberg et al., Phys. Rev. Lett. {\bf 61}, 818 (1988).
\bibitem{Lopez} J. Lopez et al., Phys. Let. B {\bf 219}, 215 (1989).
\bibitem{Colonna} M. Colonna et al., Nucl. Phys. A {\bf 613}, 165 (1997).
\bibitem{Borderie_JPG} B. Borderie, Jour. Phys. G {\bf 28}, 217 (2002).
\bibitem{Dorso} C. O. Dorso, V. C. Latora, and A. Bonasera, Phys. Rev. C {\bf 60}, 034606 (1999).
\bibitem{Campi} X. Campi, Phys. Lett B {\bf 208}, 351 (1988).
\bibitem{Elliott} J. B. Elliott et al., Phys. Rev. Lett. {\bf 88}, 042701 (2002).
\bibitem{Ma} Y. G. Ma, Phys. Rev. Lett. {\bf 83}, 3617 (1999).
\bibitem{Mallik16} S. Mallik, G. Chaudhuri, P. Das and S. Das Gupta, Phys. Rev. C {\bf 95}, 061601 (2017)(R).
\bibitem{Das} C. B. Das, S. Das Gupta, W. G. Lynch, A.Z. Mekjian and M. B. Tsang, Phys. Rep. {\bf 406}, 1 (2005).
\bibitem{Lin1} W. Lin, P. Ren, H. Zheng, X. Liu, M. Huang, R. Wada and G. Qu, Phys. Rev. C {\bf 97}, 054615 (2018).
\bibitem{Lin2} W. Lin, P. Ren, H. Zheng, X. Liu, M. Huang, K. Yang, G. Qu, R. Wada, Phys. Rev. C {\bf 99}, 054616 (2019).
\bibitem{Liu} H. L. Liu, Y. G. Ma and D. Q. Fang, Phys. Rev. C {\bf 99}, 054614 (2019).
\bibitem{Mallik20} S. Das Gupta, S. Mallik and G. Chaudhuri, Phys. Rev. C {\bf 97}, 044605 (2018).
\bibitem{Lin3} W. Lin, P. Ren, X. Liu, H. Zheng,  M. Huang, G. Qu, R. Wada, , Jour. Phys. G {\bf 48}, 085103 (2021).
\bibitem{Bakeer} R. Bakeer, W. Awad, H.R. Jaqam, Jour. Phys. G {\bf 28}, 217 (2002).
\bibitem{Wada} R. Wada et al., Phys. Rev. C {\bf 99}, 024616 (2019).
\bibitem{Bao-an-li2} Bao-An Li and Wolf-Udo Schroder, {\it Isospin Physics in Heavy-Ion Collisions at Intermediate Energies}, New York: Nova Science Pub. Inc. (2001).
\bibitem{Bao-an-li1} Bao-An Li, Lie-Wen Chen and Che Ming Ko, Phys. Rep. {\bf 464}, 113, (2008).
\bibitem{Mallik25} S. Mallik, Phys. Rev. C {\bf 107}, 054605 (2023).
\bibitem{Margueron2018} J. Margueron, R. Hoffmann Casali and F. Gulminelli, Phys. Rev. C {\bf 97}, 025805 (2018).
\bibitem{Sly5} E. Chabanat, P. Bonche, P. Haensel, J. Meyer, and R. Schaeffer,
Nucl. Phys. A {\bf 635}, 231 (1997).
%Model
\bibitem{Chase} K.C. Chase and A. Z. Mekjian, Phys. Rev. C {\bf 52}, R2339 (1995).
\bibitem{Mallik_NuclAstro1} S. Mallik and F. Gulminelli, Phys. Rev. C {\bf 103}, 015803 (2021).
\bibitem{Gulminelli2015} F. Gulminelli and Ad.R. Raduta, Phys. Rev. C {\bf 92} (2015) 055803.
\bibitem{papa}P. Papakonstantinou, J. Margueron, F. Gulminelli, A.R. Raduta, Phys. Rev. C {\bf 88}, 045805 (2013).
\bibitem{Ducoin1} C. Ducoin, J. Margueron, C. Providencia and I. Vidana, Phys. Rev. C {\bf 83}, 045810 (2011).
\bibitem{LS} J. M. Lattimer and F. Douglas Swesty, Nucl. Phys. {\bf A535}, 331 (1991).
\bibitem{Carreau2019} T.Carreau, F. Gulminelli,  J.Margueron, Eur. Phys. J. A {\bf 55}, 188 (2019).
\bibitem{Carreau2019a}T. Carreau, F. Gulminelli, and J. Margueron, Phys. Rev. C {\bf 100}, 055803 (2019).
%Results
\bibitem{Frankland} J. D. Frankland et al., Nucl. Phys. A 649, 940 (2001).
\bibitem{Xu} H. S. Xu et al., Phys. Rev. Lett. {\bf 85}, 716 (2000).
\bibitem{Mallik11} S. Mallik, G. Chaudhuri and S. Das Gupta, Phys. Rev. C {\bf 91}, 044614 (2015).
\bibitem{Mallik23} S. Mallik and G. Chaudhuri,Nucl. Phys. A {\bf 1002}, 121948 (2020).
\bibitem{Mallik19} P. Das, S. Mallik and G. Chaudhuri, Phys. Let. B {\bf 783}, 364 (2018).
\bibitem{Lalazissis} G. A. Lalazissis, J. König, and P. Ring, Phys. Rev. C {\bf 55}, 540 (1997).
\bibitem{Nguyen} Nguyen Van Giai and H. Sagawa, Phys. Lett.B {\bf 106}, 379 (1981).
\bibitem{Ducoin} C. Ducoin, Ph. Chomaz and F. Gulminelli, Nucl. Phys. A {\bf 771}, 68 (2006).
\bibitem{Peaslee} G. F. Peaslee et al., Phys. Rev. {\bf C 49}, 2271 (1994)(R).
\bibitem{Ogilvie} C. A. Ogilvie et al., Phys. Rev. Lett. {\bf  67}, 1214 (1991).
\bibitem{Baym} G. Baym, H. A. Bethe and C. J. Pethick, Nucl. Phys. A {\bf 175}, 225 (1971).
\end{thebibliography}
\end{document}